\documentclass[prb,twocolumn,superscriptaddress]{revtex4}
\usepackage{mathrsfs}
\usepackage{amsfonts}
\usepackage{amssymb}
\usepackage{epsfig}
\usepackage{amsbsy}
\usepackage{amsmath}
\usepackage{graphicx}

\begin{document}

\title{Abelian and non-abelian anyons in integer quantum anomalous Hall
effect and topological phase transitions via superconducting
proximity effect}

\author{ Xuele Liu} \affiliation{Institute of Theoretical Physics, Chinese Academy of
Sciences, P.O. Box 2735, Beijing 100190, China}\affiliation{
Department of Physics, Oklahoma State University, Stillwater,
Oklahoma 74078, USA}
\author{Ziqiang Wang}
\affiliation{Department of Physics, Boston College, Chestnut
Hill, MA 02467, USA} \author{X.C. Xie}\affiliation{International
Center for Quantum Materials, Peking University, Beijing 100871,
China }\affiliation{ Department of Physics, Oklahoma State
University, Stillwater, Oklahoma 74078, USA}
\author{Yue Yu}\altaffiliation{Correspondences send to yyu@itp.ac.cn}
\affiliation{Institute of Theoretical Physics, Chinese Academy of
Sciences, P.O. Box 2735, Beijing 100190, China}
\date{\today}

\begin{abstract}
We study the quantum anomalous Hall effect  described by a class
of two-component Haldane models on square lattices. We show that
the latter can be transformed into a pseudospin triplet
$p+ip$-wave paired superfluid. In the long wave length limit, the
ground state wave function is described by Halperin's $(1,1,-1)$
state of neutral fermions analogous to the double layer quantum
Hall effect. The vortex excitations are charge $e/2$ {\it abelian
anyons} which carry a neutral Dirac fermion zero mode. The
superconducting proximity effect induces `tunneling' between
`layers' which leads to topological phase transitions whereby the
Dirac fermion zero mode fractionalizes and Majorana fermions
emerge in the edge states. The charge $e/2$ vortex excitation
carrying a Majorana zero mode is a non-abelian anyon. The
proximity effect can also drive a conventional insulator into a
quantum anomalous Hall effect state with a Majorana edge mode and
the non-abelian vortex excitations.
\end{abstract}

 \maketitle

\section{Introduction}

The discovery of the quantum Hall effect (QHE) \cite{vonk} opened
an era for studying topological quantum phases \cite{TKNN}. Some
twenty years ago, Haldane \cite{haldane} proposed the quantum
anomalous Hall effect (QAHE) for electrons on a two-dimensional
lattice.  This is generalized to the time reversal invariant
topological insulators\cite{rv} in two dimensions
\cite{KM1,KM2,bandinv,exp1} and three dimensions
\cite{MB,FKM,Roy,FK1,exp2}. With the band inversion in these
system, it raises the hope for realizing the QAHE in a two-band
model of two-dimensional magnetic insulators \cite{qzw}.
Candidate materials for this effect, HgTe doped with Mn
\cite{liu} and a tetradymite semiconductors doped with transition
metal elements \cite{df}, have been predicted. Other proposals
are also made in condensed matter systems
recently\cite{other1,other2}. The QAHE has also been proposed for
cold atom systems \cite{wangzd,LLWS,wu}.

Another advance is the search for topological phases with
non-abelian anyons \cite{MR} that have potential applications for
quantum computing \cite{kitaev}. In addition to the $\nu=5/2$
fractional QHE, it has been shown theoretically that the
superconducting  proximity effect on the surface state of
topological insulator \cite{FK} and on semiconductors with strong
spin-orbit coupling and Zeeman splitting \cite{sau} provides a
new avenue for generating the Majorana zero mode and non-abelian
vortex excitations.

In this work, we study the QAHE for a class of two-component
Haldane models on a square lattice. The physical degrees of
freedom represented by the components depend on the microscopic
details: the real spins of electrons, band indices \cite{qzw},
the top-bottom surface states of three-dimensional topological
insulator \cite{df}, as well as the sublattice indices as
exemplified below. We show that, in a pseudospin representation,
the QAHE system can be transformed into a chiral $p+ip$-wave
pairing state involving both pseudospin components. The ground
state wave function is given by a determinant of the pairing
functions whose long wave length limit is a charge neutral
Halperin $(1,1,-1)$ state analogous to the double layer QHE
\cite{halp}. There are abelian anyon excitations with charge
$e/2$, despite that the present model describes an {\it integer}
QAHE system.

When s-wave superconductor develops in the QAHE system due to the
proximity effect, tunneling between the different isospin
components takes place. We find that the system displays
continuous transitions between topological phases with abelian
and non-abelian anyon excitations. Specifically, a topological
phase transition from the Hall conductance $2e^2/h$ to ${e^2}/h$
happens for sufficiently strong proximity induced pairing because
one of the pseudospin components is driven to a strong pairing
state analogous to that in the $\nu=1/2$ double-layer fractional
QHE discussed by Ho \cite{Ho} and Read and Green \cite{read}. The
remaining pseudospin component is in the weak pairing state
described by a Moore-Read Pfaffian \cite{MR}. A Majorana zero
mode appears in the edge state and the vortex excitation carrying
this Majorana mode is a non-abelian anyon. Interestingly, the
model also exhibits a topological trivial phase without the QAHE
when the triplet $p+ip$-wave pairing is in a {\it topologically
unprotected} weak pairing state. We find that the proximity
effect can drive one of the pseudospin components into a
topologically protected weak pairing state with quantized Hall
conductance $-{e^2}/h$, edge Majorana zero mode, and non-abelian
anyon vortex excitations (see also Ref.[\onlinecite{QHZ}]).

This paper was organized as follows:

\section{Models}

The Hamiltonian of the two-component model in the
lattice momentum space is given by
\begin{eqnarray}
H_0&=&\sum_{\bf k}[(p_x+ip_y)c^\dag_{a\bf k} c_{b\bf
k}+h.c.\label{kh}\\
&+&h_z({\bf k})(c^\dag_{a\bf k}c_{a\bf k}-c^\dag_{b\bf k}c_{b\bf
k}) +h_0({\bf k})(c^\dag_{a\bf k}c_{a\bf k}+c^\dag_{b\bf
k}c_{b\bf k})]\nonumber
\end{eqnarray}
where $c_{a (b)\bf k}$ annihilate an electron of component $a(b)$
with momentum ${\bf k}$ and $h_0({\bf k})$ is the dispersion due
to hopping among electrons of the same component. The physical
origin of the terms proportional to $p_x+ip_y$ and $h_z$ depends
on the system of interest. If $(a,b)$ label the electron spin,
$p_x+ip_y\to k_x+ik_y$ arises from Rashba spin-orbit coupling
\cite{qzw} and $h_z$ is the magnetization. If $(a,b)$ are the
orbital indices, $p_x+ip_y$ describes the orbital hybridyzation
\cite{qzw,liu} and $h_z$ is the crystal field splitting. For the
doped tetradymite semiconductors, they constitute the spin-orbit
coupling associated with the three-dimensional topological
insulator \cite{df}.

\subsection{A Generalized Haldane Model in Square Lattice}

\begin{figure}[htb]
\begin{center}
\includegraphics[width=8.5cm]{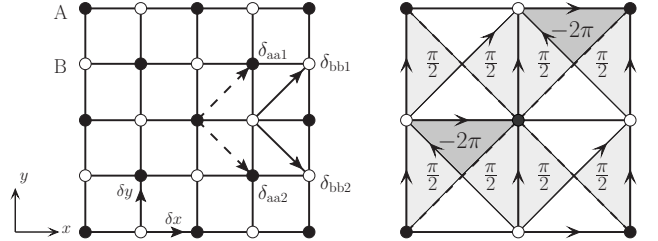}
\end{center}
\caption{\label{fig:Fig. 1} The square lattice model for
Hamiltonian (1) Left: The two-sublattice and hopping. Right: The
flux distribution. Hopping along arrowed vertical links generates
a phase $\pi$ and arrowed horizontal links a phase $\pi/2$. A net
flux of $-2\pi$ ($\pi/2$) is accumulated for the dark (light)
grey \emph{triangular blocks}. The rest of the hopping is real.}
\end{figure}

In Fig.~\ref{fig:Fig. 1}, we give an explicit realization of the
Hamiltonian (\ref{kh}) for spinless fermions on a square lattice
where $(a,b)$ label the A and B sublattices. This turns out to be
a modified Haldane's model where the complex hopping induces
$\pm\pi$ staggered plaquette flux with the link phase
distribution shown in Fig.\ref{fig:Fig. 1} (right panel). The
corresponding Halimtonian with this setup is given by
\begin{eqnarray}
H_0&=&-t\sum_{i_a}[c_{i_a+\delta_x}^\dag
c_{i_a}+ic_{i_a+\delta_y}^\dag c_{i_a}+h.c.]\nonumber\\
&+&t\sum_{i_b}[c_{i_b+\delta_x}^\dag
c_{i_b}-ic_{i_b+\delta_y}^\dag c_{i_b}+h.c.]\nonumber\\
&-&t'\sum_{i_a}[c_{i_a+\delta_{aa1}}^\dag
c_{i_a}+c_{i_a+\delta_{aa2}}^\dag c_{i_a}+h.c.]\label{oh}\\
&+&t'\sum_{i_b}[c_{i_b+\delta_{bb1}}^\dag
c_{i_b}+c_{i_b+\delta_{bb2}}^\dag c_{i_b}+h.c.]\nonumber\\
&-&M\sum_{i}(c^\dag_{i_a}c_{i_a}-c^\dag_{i_b}c_{i_b})\nonumber
\end{eqnarray}
where $c^\dag_{i_{a,b}}$ is the creation operator of the electron
at the site $i_{a,b}$ on the $A$ or $B$ sublattice. The
coordinates and vectors are figured in the left panel of Fig.
\ref{fig:Fig. 1}. The lattice spacing $a$ is set to be unit. $t$
and $t'$ are the nearest neighbor hopping amplitude and the next
nearest neighbor's. $\pm M$ is the on-site energy on $i_a$ and
$i_b$, respectively. It can also be realized in cold atom context
with a simpler staggered flux while the atom orbital is different
on the $A$ and $B$ sublattice \cite{LLWS}.

Making the Fourier transformation, the Hamiltonian is exactly
given by Eq. (\ref{kh}) with $p_x+ip_y=-2t(\sin k_y +i \sin
k_x)$, $h_z({\bf k})=-M-4t^\prime\cos k_x\cos k_y$, and $h_0=0$,
where $t$ and $t^\prime$ are the hopping amplitudes between
nearest and next nearest neighbors indicated in Fig.\ref{fig:Fig.
1} and $\pm M$ is the on-site energy for the A and B sublattices.
Note that the `magnetization' contains $\cos k_x\cos k_y$ and is
different from that used in \cite{qzw}. A similar model can also
be realized in cold atom systems \cite{LLWS}.  For $h_0=0$, there
is an important $O(2)$ symmetry associated with the $U(1)$
particle number conservation and a $\mathbb{Z}_2$ under
$c^\dag_{a,\bf k}\to ic_{b,-\bf k}$, i.e., a $\cal{C\cdot S}$
(particle-hole$~\cdot~$sublattice) symmetry. If $h_0\ne 0$ but
can be adiabatically driven to zero, we think the results of this
work are also valid.

We now study Hamiltonian (\ref{kh}) in the A-B sublattice where
 $\bf k$ is confined to the {\it reduced} first
Brillouin zone bounded by $k_x\pm k_y=\pm \pi$ due to the A-B
sublattices. Our results can be extended directly to other
relevant cases discussed above with lattice translation symmetry.

The eigen-energy of (\ref{kh}) is given by
$E_0=\pm\sqrt{h_z^2+|p|^2}$. There are two independent Dirac
points at $(0,0)$ and $(\pi,0)$ in the reduced zone. When an
extended s-wave pairing is induced by proximity effect \cite{FK}
on the QAHE system, the total Hamiltonian is given by
$H=H_0+H_{sc}$, where
\begin{equation}
H_{sc}=\sum_{\bf k} f({\bf k})(c_{a-\bf k}c_{b \bf
k}+c^\dag_{b\bf k} c^\dag_{a-\bf k}), \label{hsc}
\end{equation}
and $f({\bf k})=2\Delta(\cos k_x+\cos k_y)$. The eigen-energy of
the total Hamiltonian is given by $ E=\pm\sqrt{(h_z\pm
f)^2+|p|^2}$. The energy gap of the QAHE at the Dirac point
$(0,0)$ closes when $\Delta=\frac{1}4(M+4t^\prime)$, whereas the
QAHE gap at $(\pi,0)$ is unperturbed since $f({\bf k})=0$ for
$k_x\pm k_y=\pm\pi$.

To unveil the ground state wavefunction and the topological
properties of Eq. (\ref{kh}), we introduce a pseudospin
representation by mixing the electron and hole of different
component (which in fact indicates a particle-hole transformation
of $b$-component)
\begin{eqnarray}
c_{\uparrow\bf k}=(c_{a\bf k}+c^\dag_{b-\bf
k})/{\sqrt2},~~c_{\downarrow\bf k}=i(c_{a\bf k}-c^\dag_{b-\bf
k})/{\sqrt2}.\label{trans}
\end{eqnarray}
Under this unitary transformation, the Hamiltonian in terms of
fermions carrying the pseudospin becomes
\begin{eqnarray}
H_0=\sum_{{\bf k};s=\uparrow,\downarrow} [h_zc^\dag_{s \bf
k}c_{s\bf k}-((p_x-ip_y) c_{s\bf k}c_{s-\bf k}+h.c.)/2].
\label{ps}
\end{eqnarray}
Both the pseudospin-$\uparrow$ and $\downarrow$ fermions, having
a band dispersion  $h_z({\bf k})$, are in the $p+ip$-wave paired
states. Hamiltonian (\ref{ps}) is closely related to the
$\nu=1/2$ double-layer fractional QHE if we identify the
pseudospins with the even/odd states of the isospin (layer index)
in the context of triplet chiral $p$-wave pairing \cite{Ho,read}.

\subsection{Winding Numbers}

\begin{figure}[htb]
\begin{center}
\includegraphics[width=7cm]{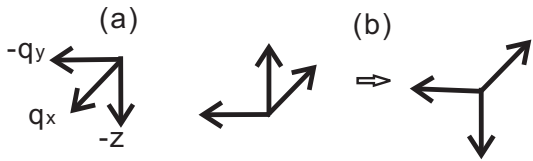}
\end{center}
\caption{\label{fig:Fig. 2}: The frames and frame change from
$M<4t'$ to $M>4t'$. (a) $(-2tq_y,2tq_x,-M-4t')$ at $(0,0)$. The
frame is not changed and is the right. (b)
 $(-2tq_y,-2q_x,-M+4t')$ at $(\pi,0)$.
 The frame is changed from the right for $M<4t'$ to the left for $M>4t'$. }
\end{figure}

We now calculate the winding number, which describes the mapping
from the reduced zone (a torus) to a target sphere specified by
the unit vector ${\bf n}=\frac{(p_x,p_y,h_z)}{E_0}$. The winding
number is given by $C=C_\uparrow+C_\downarrow$ with
$C_s=\frac{1}{4\pi}\int d^2k{\bf n}\cdot\partial_{k_x}{\bf
n}\times\partial_{k_y}{\bf n}$ in the continuum limit. A direct
calculation yields $C_\uparrow=C_\downarrow=1$ and $C=2$ which
reflects the fact that the whole first Brillouin zone covers
twice of the sphere with such a map ${\bf n}$. This result can be
understood intuitively by considering the vector components of
${\bf n}$ near the two Dirac points $(0,0)$ and $(\pi,0)$. For a
small deviation ${\bf q}\sim 0$, they are
$(-2tq_y,2tq_x,-M-4t')/E_0$ near $(0,0)$ and
$(-2tq_y,-2q_x,-M+4t')/E_0$ near $(\pi,0)$ (See Fig.
\ref{fig:Fig. 2}). Therefore, the Dirac points are mapped to the
north and south poles which are covered once. A semi-sphere
including a pole contributes $\pm1/2$ to the winding number
depending on the pole's frame. When $M<4t'$, both poles are in
the right hand frame and each semi-sphere contributes 1/2 which
gives $C_{\uparrow,\downarrow}=1$ and $C=2$. Thus, this QAHE has
Hall conductance $2\frac{e^2}h$.

\section{Ground states and anyonic excitations}

\subsection{The (1,1,-1)-state and abelian anyons}

The paired state has BCS coherence factors $|u_{s,\bf
k}|^2=\frac{1}2(1+\frac{h_z}{E_0}), |v_{s,\bf
k}|^2=\frac{1}2(1-\frac{h_z}{E_0})$ and pairing functions
\begin{equation}
g_s({\bf k})={v_{s,\bf k}}/{u_{s,\bf k}}=g({\bf
k})=-({E_0-h_z})/p. \label{pairingfunction}
\end{equation}
For $M<4t'$, the effective chemical potential at the Dirac point
$(\pi,0)$ $\mu_{(\pi,0)}=-h_{z,(\pi,0)}=M-4t'<0$ such that
$u_{s}\sim 1$ and $g_s\sim v_s\sim q_y+iq_x$. This is in the
 strong pairing regime \cite{read} and can be
thought as the `infinity' point in the continuum theory. On the
other hand, the effective chemical potential at the Dirac point
$(0,0)$ $\mu_{(0,0)}=-h_{z;(0,0)}=M+4t'>0$ which leads to
$v_s\sim 1$ and $g_s\sim 1/u_{s}\sim 1/ (q_y-iq_x)$. This
singular pairing function in the long wavelength limit is the
hallmark of the topologically nontrivial $p+ip$ weak pairing
phase \cite{read}. Therefore, the long distance, low energy
physics is determined by the weak pairing of the qusiparticles
carrying both pseudospins near the Dirac point $(0,0)$. The
ground state is given by
\begin{eqnarray}
|G_s\rangle\propto\exp[\frac{1}2\sum_{\bf k}g({\bf
k})(c^\dag_{\uparrow\bf k}c^\dag_{\uparrow\bf
-k}+c^\dag_{\downarrow\bf k}c^\dag_{\downarrow\bf -k})]|0_s\rangle,
\label{gs}
\end{eqnarray}
where $\vert 0_s\rangle$ is the vacuum for the fermions carrying
the pseudospin,  $c_{s{\bf k}}\vert 0_s\rangle=0$. From the
transformation (\ref{trans}), it is clear that this vacuum is
empty of the $a$-electrons but filled with the $b$-electrons. The
ground state in Eq.~(\ref{gs}) resembles the neutral part of the
$(3,3,1)$-state in $\nu=1/2$ double layer fractional QHE
\cite{halp,read}. Going back to the original $a$-$b$ component
electrons, the ground state wave function of Eq. (\ref{kh}) is of
the form of a {\sl determinant of the pairing function} $g({\bf
r})$, the Fourier image of $g({\bf k})$,
$$
\Psi(\{{\bf r}_i^a,{\bf r}_i^b\})=
\langle 0|\prod_i c_{a}({\bf r}_i^a)c_{b}
({\bf r}_i^b)|G\rangle\propto {\det}[g({\bf r}^a_i-{\bf r}^b_j)],
$$
where ${\bf r}_i^{a,b}$ are the coordinates of the two-component
electrons in the ground state $|G\rangle$ out of the vacuum
$c_{a(b)k}\vert0\rangle=0$. The ground state $|G\rangle$ and the
vacuum $\vert0\rangle$ of the original Hamiltonian (\ref{kh}) are
related to $|G_s\rangle$ and $\vert0_s\rangle$ in the pseudospin
representation by $|G_s\rangle=U|G\rangle$,
$\vert0_s\rangle=U\vert0\rangle$ where $U$ is the unitary operator
for the particle-hole transformation in the $b$-component indicated
in Eq. (\ref{trans}). In the long wavelength limit, the weak pairing near $(0,0)$
implies $g({\bf r}^a_i-{\bf r}^b_j)\sim\frac{1}{z^a_i-z^b_j}$. Using Cauchy's determinant identity, we find
\begin{eqnarray}
\Psi(\{{\bf r}_i^a,{\bf r}_i^b\})\!\sim\!
\prod_{i<j}(z^a_i-z^a_j)\prod_{k<l}(z^b_k-z^b_l)
\prod_{sr}(z^a_s-z^b_r)^{-1}.\nonumber
\end{eqnarray}
This is exactly the
charge neutral part of the (3,3,1), i.e. a $(1,1,-1)$-state.

The above result allows us to describe localized charge
excitations in the filled bands which are carried by the finite
${\bf k}$ states. A Laughlin-type quasihole excitation is a good
approximation of a real hole with charge $e$ because the number
of extended states in the QAHE system is the same as that in
a band insulator. Thus, the minimal charge $e/2$ vortex pair
excitations located at $\eta_1$ and $\eta_2$ are similar to that
in the (3,3,1) state, i.e.,
$$
\Psi_v(\eta_1,\eta_2)\sim{\det}\{[(z^a_i-\eta_1)(z^b_j-\eta_2)
+(1\leftrightarrow 2)]g({\bf r}^a_i-{\bf r}^b_j)\}.
$$
We thus come to the conclusion that the vortex excitations in
this integer QAHE system are charge $e/2$ abelian anyons carrying
a neutral Dirac fermion zero mode. This fractional charge can
also be understood from the ground state degeneracy. In Eq.
(\ref{ps}), the $O(2)$ symmetry is generated by the pseudospin
$S_y=-i\sum_{\bf k}(c^\dag_{\uparrow \bf k} c_{\downarrow\bf
k}-c^\dag_{\downarrow\bf k} c_{\uparrow\bf k} )/2$ and
$c_{\uparrow \bf k}\leftrightarrow c_{\downarrow\bf k}$. Thus,
the ground state is four-fold degenerate and the vortex
excitations carry $S_y=1/4$ \cite{read}. In terms of the original
QAHE system, $2S_y$ corresponds to the charge operator $N_a+N_b$
due to the partial particle-hole transformation. Thus, the vortex
carries charge $e/2$, which is different from $e/4$ in the
(3,3,1)-state. The vortex pairs have an excitation energy on the
order of $e^2/(4|\eta_1-\eta_2|)$ and are thus stable when the
vortices are well separated by the distance $l>e^2/[8(M+4t')]$.
When the vortices are close enough so that their energy is higher
than the band gap, they fuse into a real hole.

\subsection{ Weak-strong pairing phase transition and non-abelian
anyons}

 In the pseudospin representation, the pairing due to the
proximity effect in Eq.~(\ref{hsc}) plays the role of a magnetic
field and $h_z$ in Eq.~(\ref{ps}) splits into
$h_{z;\uparrow,\downarrow}=h_z\mp f({\bf k})$ for
$c_{\uparrow,\downarrow}$, respectively. In the double-layer
language, this $f({\bf k})$ amounts to interlayer tunneling that
splits the symmetric and antisymmetric combinations. The
antisymmetric component can be driven into a strong pairing
state, leaving only the symmetric one governing the physics at
low energy and long wavelength \cite{Ho}. The winding number
associated with the pseudospin-up (-down) component is defined by
the unit vector ${\bf
n}_{\uparrow(\downarrow)}=(p_x,p_y,h_{z;\uparrow(\downarrow),\bf
k})/E$. Since $f({\bf k})=0$ at $(\pi,0)$, ${\bf
n}_{\uparrow,\downarrow}= {\bf n}$ at this Dirac point. Near the
Dirac point $(0,0)$, ${\bf
n}_{\uparrow,\downarrow}=(-2tq_y,2tq_x,-M-4t'\mp 4\Delta)/E$. For
$M<4t'$, the energy gap at $(0,0)$ closes at
$\Delta=\Delta_0=\frac{1}4(M+4t')$. When $\Delta>\Delta_0$, ${\bf
n}_\downarrow$ at $(0,0)$ changes from the right frame to the
left frame while ${\bf n}_\uparrow$ remains in the right frame.
Therefore, $C_\uparrow=1$ while the pseudospin-down component
becomes topological trivial with $C_\downarrow=0$. This can also
be seen from the effective chemical potential of the components;
only $\mu_{\uparrow,(0,0)}=-h_{z,\uparrow;(0,0)}>0$ satisfies the
weak pairing condition, while the other three are in the strong
pairing regime, i.e.
$\mu_{\downarrow;(0,0)}=-h_{z,\downarrow;(0,0)}<0$ and
$\mu_{\uparrow,\downarrow;(\pi,0)}=-h_{z,\uparrow,\downarrow;(\pi,0)}<0$.
Thus, the ground state wave function is given by $ \Psi(\{{\bf
r}_i^a,{\bf r}_i^b\})\sim{\rm Pf}[g_\uparrow({\bf r}_i-{\bf
r}_j)], $ which is identical to the charge neutral Moore-Read
Pfaffian in the long wavelength limit. The ground state is
three-fold degenerate \cite{read}. A pair of vortices are the
same as the Moore-Read non-abelian anyons \cite{MR}
$$
\Psi_v(\eta_1,\eta_2)\sim{\rm
Pf}\{[(z_i-\eta_1)(z_j-\eta_2)+(1\leftrightarrow2)]
g_\uparrow({\bf r}_i-{\bf r}_j)\}.
$$
However, each vortex again carries charge $e/2$. The safe
distance between the vortices is also determined by the band gap.

\begin{figure}[htb]
\begin{center}
\includegraphics[width=8.5cm]{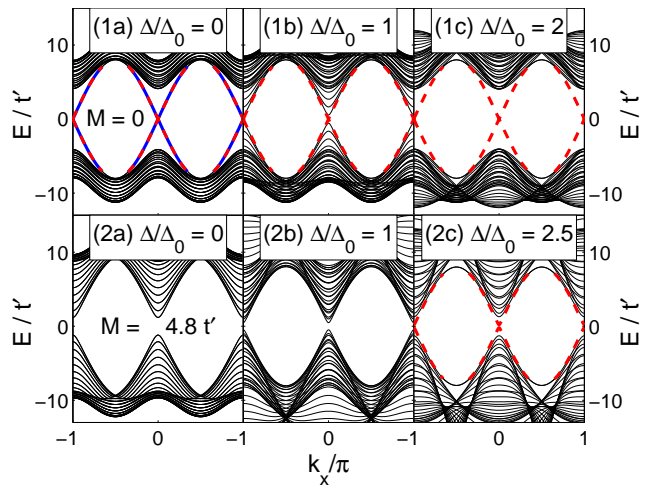}
\end{center}
\caption{\label{fig:Fig. 3}(color online) The energy dispersion
$E$ along $k_{x}$ showing the evolution of the edge states with
proximity induced pairing $\Delta$. Top panel starts with (1a)
the QAHE state ($M<4t^\prime$) while the bottom panel starts with
(1b) the insulating state ($M>4t^\prime$) when $\Delta=0$. The
gap closes at $\Delta=\Delta_0$ (1b and 2b). The chiral edge
states located at the $y=0$ and $y=L$ boundaries are marked by
red dashed lines and, in the case of degeneracy, the blue solid
lines. $t=4t'$.}
\end{figure}

\subsection {Insulator-QAHE transition and non-abelian anyons}

Interestingly, for $M>4t'$, the model $H_0$ with the proximity
effect is topologically trivial with $C_{\uparrow,\downarrow}=0$,
since the poles at $(\pi,0)$ and $(0,0)$ are in the opposite
frames (See Fig. \ref{fig:Fig. 2}(b)). In this case $v_s\sim 1$,
but $u_s\sim q_y-iq_x$ near $(0,0)$ while $u_s\sim q_y+iq_x$ near
$(\pi,0)$. As a result, the quasiparticle local wave functions
near the Dirac points are the holomorphic $(1,1,-1)$-state for
$(0,0)$ and the anti-holomorphic $(1,1,-1)$-state for $(\pi,0)$.
They contribute to the Hall conductance with the same magnitude
but opposite signs and lead to an insulating state.

In the presence of proximity induced pairing, the insulating gap
closes at $\Delta=\Delta_0$. When $\Delta>\Delta_0$, the
pseudospin-up component is pushed into the strong pairing regime
near (0,0), i.e. $\mu_{\uparrow,(0,0)}<0$, while all others
remain unchanged. Thus the low energy physics is dominated by the
Pfaffian of the up-spin component near $(\pi,0)$ with winding
numbers $C_\uparrow=-1$ and $C_\downarrow=0$. This is therefore a
transition from an insulator to a non-abelian QAHE state with
Hall conductance $-e^2/h$ where vortex excitations are
non-abelian anyons \cite{QHZ}.

\section{Edge states and Majorana fermion modes}

 Next, we
present explicit calculations of the edge state spectrum of the
total Hamiltonian $H$ under an open boundary condition along the
$y$-direction of a strip. The existence of the nontrivial
topological invariants implies stable gapless edge modes
separated from the gapped bulk excitations. For a small $M<4t'$,
say $M=0$, gapless chiral edge states appear across $k_x=0$ as
shown in Fig.~\ref{fig:Fig. 3}. There are indeed two degenerate
chiral edge states in this $(1,1,-1)$ abelian QAHE state
(Fig.~\ref{fig:Fig. 3}(1a)). The gap closing is demonstrated in
Fig. \ref{fig:Fig. 3}(1b). After the proximity induced
topological phase transition, a single edge mode survives (Fig.
\ref{fig:Fig. 3}(1c)) in the non-abelian QAHE state. The lower
panel of Fig.~\ref{fig:Fig. 3} starts with the non-topological
insulator at $M=4.8t'$ where edge states are absent. The
proximity effect causes the topological gap closing transition
(Fig.~\ref{fig:Fig. 3}(2b)) and the emergence of a single edge
state in non-abelian QAHE state (Fig.~\ref{fig:Fig. 3}(2c)).

\begin{figure}[htb]
\begin{center}
\includegraphics[width=8.5cm]{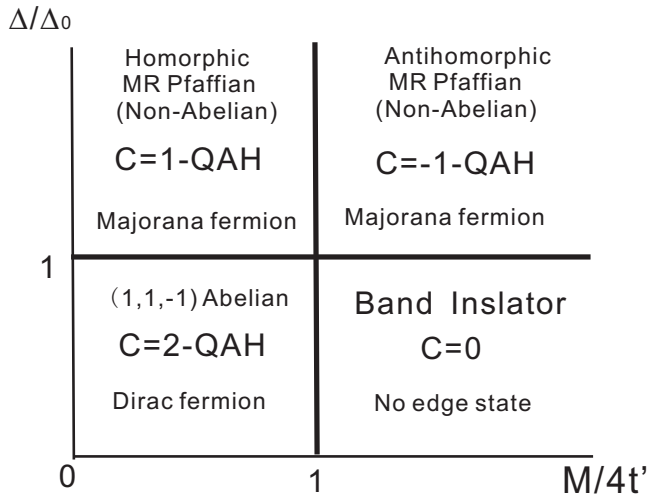}
\end{center}
\caption{\label{fig:Fig. 4} The phase diagram with the Chern
numbers, the ground state wave functions and the gapless edge
modes.}
\end{figure}

These numerical calculations confirm the analysis and results
obtained in previous sections. For $k_x=0+q_x\ll 1$, the gapless
edge modes can be written as $\psi_s(0;q_x)=u_s(0;q_x)c_{s,0;
q_x}+v_s(0;q_x)c^\dag_{s,0;-q_x}.$ At $q_x=0$, we have
$u_s(q_x)=v(q_x)=\frac{1}{\sqrt2}$ and
$\psi^\dag_s(0;0)=\psi_s(0;0)$. These are indeed the Majorana
fermion zero modes. In the (1,1,-1)-state, both
$\psi_{\uparrow,\downarrow}$ exist and combine into a complex
neutral Dirac fermion, which is consistent with the neutrality of
the bulk ground state in the long wavelength limit. For both the
holomorphic Moore-Read Pfaffian (Fig. \ref{fig:Fig. 3}(1c)) and
the anti-holomorphic Pfaffian states (Fig. \ref{fig:Fig. 3}(2c))
induced by the proximity effect, there is only one Majorana mode
on the edge, consistent with the existence of the non-abelian
anyons.

\section{Phase Diagram and Conclusions}

In Fig. \ref{fig:Fig. 4}, we summarize our results in a schematic
phase diagram for this class of two-component abelian and
nonabelian QAHE. While the interesting non-abelian phase was
confirmed after a superconductor proximity, the ${\cal C\cdot S}$
discrete symmetry leads to a new topological order with abelian
anyon even without a superconductor proximity and no interaction
between electrons are concerned. Finally, we argued that the
partial particle-hole transformation may be a general duality
between a topological insulator and a topological superconductor.

 \vspace{0.2cm}

\centerline{\bf ACKNOWLEDGEMENTS}

We thank Dung-Hai Lee for useful discussions. This work was
supported by National Natural Science Foundation of China, the
national program for basic research of MOST of China (YY,XX), the
Key Lab of Frontiers in Theoretical Physics of CAS(YY), in part
by DOE DE-FG02-99ER45747 and NSF DMR-0704545 (ZW), and DOE
DE-FG02-04ER46124(XL,XX).


\begin{thebibliography}{99}
\bibitem{vonk} K. von Klitzing, G. Dorda, and M. Pepper, Phys. Rev. Lett. {\bf 45}, 494(1980).

\bibitem{TKNN} D. J. Thouless, M. Kohmoto, M.P. Nightingale, and M. den Nijs, Phys. Rev. Lett.
{\bf 49}, 405 (1982).



\bibitem{haldane} F. D. M. Haldane, Phys. Rev. Lett. {\bf 61}, 2015(1988).

\bibitem{rv} For rewiev, see  M. Z. Hasan and C. L. Kane,
 Rev. Mod. Phys. {\bf 82}, 3045 (2010)
; X. L. Qi and S. C. Zhang, arXiv: 1008.2026 (2010).


\bibitem{KM1} C. L. Kane and E. J. Mele, Phys. Rev. Lett. {\bf
95}, 146802 (2005).

\bibitem{KM2} C. L. Kane and E. J. Mele, Phys. Rev. Lett. {\bf
95},  226801 (2005).





\bibitem{bandinv} B. A. Bernevig, T. L. Hughes, and S. C. Zhang, Science
{\bf 314}, 1757 (2006).

\bibitem{exp1} M. K\"onig, S. Wiedmann, C. Br\"une, A. Roth, H. Buhmann,
 L.  Molenkamp, X. L. Qi  and S. C.  Zhang, Science {\bf 318}, 766 (2007).

\bibitem{MB} J. E. Moore and L. Balents, Phys. Rev. B {\bf 75}, 121306 (2007).


\bibitem{FKM} L. Fu, C. L. Kane, and E. J. Mele , Phys. Rev. Lett. {\bf 98},
106803 (2007).

\bibitem{Roy} R. Roy, Phys. Rev. B {\bf 79}, 195322 (2009).

\bibitem{FK1} L. Fu, and C. L. Kane, Phys. Rev. B {\bf 76}, 045302 (2007).

\bibitem{exp2} D. Hsieh, D. Qian, L. Wray, Y. Xia, Y. S.  Hor , R. J. Cava and M. Z. Hasan,
Nature {\bf 452}, 970 (2008).



\bibitem{qzw} X. L. Qi, Y. S. Wu, and S. C. Zhang, Phys. Rev. B {\bf 74}
, 085308 (2006).

\bibitem{liu} C. X. Liu, X.-L. Qi, X. Dai, Z. Fang, and S.-C. Zhang, Phys. Rev. Lett. {\bf
101}, 146802 (2008).

\bibitem{df} R. Yu, W. Zhang, H. J. Zhang, S. C. Zhang, X. Dai, and Z. Fang
, Science {\bf 329}, 61 (2010).


\bibitem{other1} Y. Zhang and C. Zhang, arXiv:1009.1200.

\bibitem{other2} Z. Qiao, S. A. Yang, W. Feng, W.-K. Tse, J. Ding, Y. Yao, J.
Wang and Q. Niu, Phys. Rev. B {\bf 82}, 161414 (2010).


\bibitem{wangzd}
L. B. Shao, S.-L. Zhu, L. Sheng, D. Y. Xing, and Z. D. Wang,
Phys. Rev. Lett. {\bf 101}, 246810 (2008).

\bibitem{LLWS} X. J. Liu, X. Liu, C. Wu, and J. Sinova
, Phys. Rev. A {\bf 81}, 033622 (2010).

\bibitem{wu} M. Zhang, H.-H. Hung, C. Zhang, and C. Wu,
arXiv:1009.2133.

\bibitem{MR} G. Moore and N. Read, Nucl. Phys. B {\bf 360},
362 (1991).


\bibitem{kitaev} A. Y. Kitaev, Ann. Phys. (N.Y.) {\bf 303}, 2 (2003).



\bibitem{FK} L. Fu and C. L. Kane, Phys. Rev. Lett.
{\bf 100}, 096407 (2008).

\bibitem{sau} J. D. Sau, R. M. Lutchyn, S. Tewari, and S. Das Sarma, Phys. Rev. Lett. {\bf 104}, 040502 (2010).





\bibitem{halp} B. I. Halperin, Helv. Phys. Acta {\bf 56}, 783 (1983); Surf.
Sci. {\bf 305}, 1 (1994).


\bibitem{Ho} T. L. Ho, Phys. Rev. Lett. {\bf 75}, 1186 (1995).


\bibitem{read} N. Read and  D. Green, Phys. Rev. B {\bf 61}, 10267 (2000).

\bibitem{QHZ} X. L. Qi, T. L. Hughes, and S. C. Zhang,
Phys. Rev. B {\bf 82}, 184516 (2010).



\end{thebibliography}
\end{document}